\documentclass{ws-procs9x6}

\begin{document}

\title{eRHIC - Accelerator and detector design studies}

\author{B. Surrow}

\address{Massachusetts Institute of Technology \\
77 Massachusetts Avenue \\ 
Cambridge, MA 02139, USA \\ 
E-mail: surrow@mit.edu}

\maketitle

\abstracts{
 An electron-proton/ion collider facility (eRHIC) is under consideration at Brookhaven 
 National Laboratory (BNL). An overview of the accelerator and detector design concepts will
 be provided.}

\section{eRHIC accelerator design}
 The high energy, high intensity polarized electron/positron beam 
 ($5-10\,$GeV/$10\,$GeV) facility (eRHIC) which is under consideration at BNL
 will collide with the existing RHIC heavy ion ($100\,$GeV per nucleon) and 
 polarized proton beam ($50-250\,$GeV). This facility will allow to  
 significantly enhance the exploration of fundamental aspects of Quantum Chromodynamics 
 (QCD), the underlying quantum field theory of strong interactions~\cite{ref:annual-rev-erhic}. 
 A detailed report on the accelerator and interaction region (IR) design of this new collider facility
 has been completed based on studies performed jointly by BNL and MIT-Bates in collaboration with
 BINP and DESY~\cite{ref:erhic-zd0}. The main design option is based on the construction of a $10\,$GeV electron/positron
 storage ring intersecting with one of the RHIC hadron beams. The electron beam energy will be variable down to
 $5\,$GeV with minimal loss in luminosity and polarization. The electron injector system will consist of linacs
 and recirculators fed by a polarized electron source. A study has shown that an ep luminosity of $4\times 10^{32}$cm$^{-2}$s$^{-1}$
 can be achieved for the high-energy mode ($10\,$GeV on $250\,$GeV), if the electron beam facility is designed using today's 
 state-of-the-art accelerator technology without an extensive R\&D program. A robust cost model has
 been worked out. For electron-gold ion collisions ($10\,$GeV on
 $100\,$GeV$/u$), the same design results in a luminosity of $4\times 10^{30}$cm$^{-2}$s$^{-1}$. The potential to go to higher 
 luminosities at the level of $10\times 10^{32}$cm$^{-2}$s$^{-1}$ (high-energy ep mode) by increasing the electron beam intensity 
 will be explored in the future. A polarized positron beam of $10\,$GeV energy and high intensity will also be possible using the
 process of self-polarization. A possible alternative design for eRHIC has been presented
 on the basis of an energy recovery superconducting linac (ERL)~\cite{ref:erhic-zd0}. This option would be restricted to electrons only. 
 Preliminary estimates suggest that this design option could produce 
 higher luminosities at the level of $~10^{34}$cm$^{-2}$s$^{-1}$ (high-energy ep mode). Significant R\&D efforts for the 
 polarized electron source and for the energy recovery technology is required.
 The existing RHIC heavy-ion and polarized pp collider facility will require various additions such as the increase of the total 
 stored beam current by going to a 360 bunch mode. So far, only a 120 bunch mode has been accomplished. An electron cooling 
 system will be critical for the eA running mode to achieve and maintain small beam emittances. An upgrade of the beam source
 system (EBIS) is in preparation. 

 The eRHIC collider facility in comparison to other future high-energy DIS collider based efforts~\cite{ref:dis2006-dainton} 
 is unique since it would allow for the first time to collide 
 polarized electrons (positrons) on polarized protons as well as electrons (positrons) on light to heavy nuclei.  

\begin{figure}[t]
\centering
\includegraphics*[width=55mm]{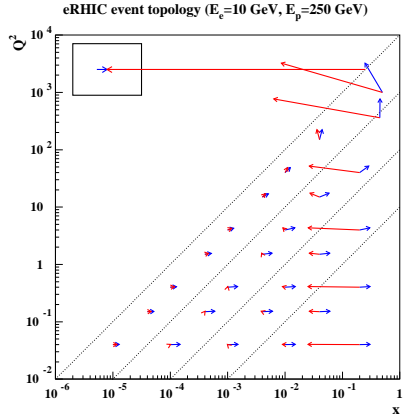}
\caption{\it Kinematic $Q^{2}$-$x$ plane for $E_{e}=10\,$GeV and $E_{p}=250\,$GeV 
showing the direction and energy of the scattered electron and current jet. The scale is taken with respect to 
the incoming electron and proton beams shown on the upper left corner.}
\label{fig:erhic-topology}
\end{figure}

\section{eRHIC detector design}

The following minimal requirements on a future eRHIC detector can be made:

\begin{itemize}
\item Measure precisely the energy and angle of the scattered electron (Kinematics of DIS reaction)
\item Measure hadronic final state (Kinematics of DIS reaction, jet studies, flavor tagging, fragmentation studies, particle ID system for heavy 
flavor physics and K/$\pi$ separation)
\item Missing transverse energy measurement (Events involving neutrinos in the final state, electro-weak physics)
\end{itemize}

In addition to those demands on a central detector, the following forward and rear detector systems are crucial:

\begin{itemize}
\item Zero-degree photon detector to control radiative corrections and measure Bremsstrahlung photons for luminosity measurements
(absolute and relative with respect to different ep spin combinations)
\item Tag electrons under small angles ($<1^{\circ}$) to study the non-perturbative and perturbative QCD transition region
\item Tagging of forward particles (Diffraction and nuclear fragments)
\end{itemize}

Figure \ref{fig:erhic-topology} shows the kinematic $Q^{2}$-$x$ plane for $E_{e}=10\,$GeV and $E_{p}=250\,$GeV 
with the direction and energy of the scattered electron and current jet for several ($Q^{2}$,\,$x$) points. In the low-$Q^{2}$/low-$x$
region both the scattered electron and current jet are predominantly found in the rear direction with energies well below
$10\,$GeV. Good electron/hadron separation is essential.
Once $x$ get larger, the current jet moves forward with larger energies and is clearly separated from the scattered
electron at low-$Q^{2}$. At high-$Q^{2}$/high-$x$ both the current jet and the scattered electron are found predominantly in the barrel
and forward direction with energies substantially larger then $10\,$GeV (electron) and $100\,$GeV (current-jet).  

\begin{figure}[ht]
\centering
\includegraphics*[width=60mm]{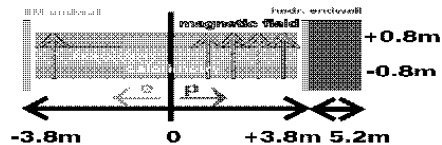}
\caption{\it Conceptual detector layout focusing on forward physics with a $7\,$m long dipole field and an
interaction region without machine elements extending from $-3.8\,$m to $+5.8\,$m.}
\label{fig2}
\end{figure}

Optimizing all of the above requirements is a challenging task. Two detector concepts have been considered so far. 
One, which focuses on the rear/forward acceptance (Figure \ref{fig:erhic-topology}) and 
thus on low-$x$/high-$x$ physics, which emerges out of the HERA-III 
detector studies~\cite{5-ref}. 
This detector concept is based on a compact system of tracking and central electromagnetic calorimetry inside a magnetic dipole field and
calorimetric end-walls outside. Forward produced charged particles are bent into the detector volume which extends the rapidity coverage 
compared to existing detectors. A side view of 
the detector arrangement is shown in Figure \ref{fig2}. The required machine element-free region amounts to roughly $\pm 5$m. This clearly 
limits the achievable luminosity
in a ring-ring configuration. 

\begin{figure}[t]
\centering
\includegraphics*[width=70mm]{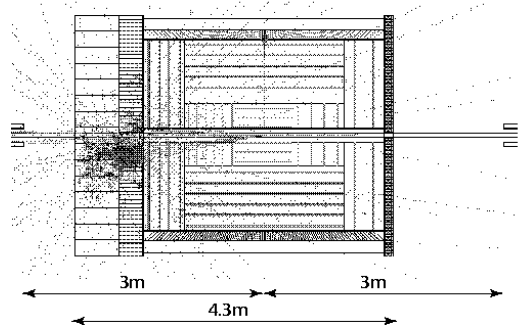}
\caption{\it Side view of the GEANT detector implementation as part of the ELECTRA simulation and reconstruction package. A deep-inelastic scattering event resulting from a LEPTO simulation is overlayed with $Q^{2}=361\,$GeV$^{2}$ and $x=0.45$ which is also drawn in Figure \ref{fig:erhic-topology}.}
\label{fig3}
\end{figure}

The second design effort focuses on a wide acceptance detector system similar to the current HERA collider experiments H1 and ZEUS to allow for
the maximum possible $Q^{2}$ range. The physics program demands high luminosity and thus focusing machine elements in a ring-ring configuration 
have to be as close as possible to the IR while preserving good central detector acceptance. This will be discussed in more detail in the next section. 
A simulation and reconstruction package called ELECTRA has been developed to design a new eRHIC detector at BNL~\cite{4-ref,44-ref}. 
Figure \ref{fig3} shows a side view
of a GEANT detector implementation of the above requirements on a central detector. The hermetic inner and outer tracking system including 
the electromagnetic
section of the barrel calorimeter is surrounded by an axial magnetic field. The forward calorimeter is subdivided into hadronic and electromagnetic 
sections. Re-using the ZEUS uranium calorimeter is under consideration. The rear and barrel electromagnetic calorimeter consists of segmented towers, e.g. a tungsten-silicon
type. This would allow a fairly compact configuration. 
The inner most double functioning dipole and quadrupole magnets are located at a distance of $\pm 3$m from the IR. An initial IR 
design assumed those inner most machine elements at $\pm 1$m. This would significantly impact the detector acceptance. 

The bunch crossing frequency amounts to roughly $30\,$MHz. This sets stringent requirements on the high-rate capability of the tracking system. 
This makes a silicon-type detector for the inner tracking system (forward and rear silicon disks together with several silicon barrel layers) 
together with several GEM-type outer tracking layers a potential choice. The forward and rear detector systems have not been considered so far. 
The design and location of those detector systems has to be worked out in close collaboration to accelerator physicists since machine magnets 
will be potentially employed as spectrometer magnets and thus determine the actual detector acceptance and ultimately the final location. It is 
understood that demands on optimizing the rear/forward detector acceptance might have consequences on the machine layout and is therefore an 
iterative process. 

\section{Considerations on the accelerator/detector interface}

The following section provides an overview of some aspects of the detector/machine interface. The specification of those items has only recently been started.

The direct synchrotron radiation has to pass through the entire IR
before hitting a rear absorber system. This requires that the geometry of the beam pipe is designed appropriately with changing shape along the longitudinal beam
direction which includes besides a simulation of the mechanical stress also the simulation of a cooling system of the inner beam pipe. The beam pipe design has
to include in addition the requirement to maximize the detector acceptance in the rear and forward direction. Furthermore the amount of dead material has
to be minimized in particular to limit multiple scattering (track reconstruction) and energy loss for particles under shallow angles (energy reconstruction). The distribution
of backscattered synchrotron radiation into the actual detector volume has to be carefully evaluated. An installation of a collimator system has to be worked out. Those
items have been started in close contact to previous experience at HERA~\cite{1-ref}.

The demand of a high luminosity ep/eA collider facility requires the installaton of focusing machine elements
as close as possible to the central detector. An IR design with machine elements at $\pm 1\,$m to the
IR would significantly limit the achievable detector acceptance. A new
scheme provides a machine-element free region of $\pm 3\,$m at the expense of approximately
half the luminosity~\cite{2-ref}. This concept is based on dipole windings as part of the detector solenoid to also achieve
an early bend of the electron beam. A linac-ring option would not be limited by beam-beam 
effects compared to a ring-ring configuration. Even larger luminosities could be achieved with a machine-element free region
of approximatley $\pm 5\,$m. 

The need for acceptance of scattered electrons beyond the central detector acceptance is driven by the need for luminosity
measurements through ep/eA Bremsstrahlung and photo-production physics. Besides that a calorimeter setup to tag radiated photons 
from initial-state radiation and Bremsstrahlung will be necessary. The scattered electrons will pass through the machine elements 
and leave the beam pipe through special exist windows. The simulation of various small-angle calorimeter setups has been started.
This will require a close collaboration with the eRHIC machine design efforts to aim for an optimal detector setup.

The forward tagging system beyond the central detector will play a crucial role in diffractive ep/eA physics. 
The design of a forward tagger system based on forward calorimetry and Roman pot stations is foreseen. Charged particles will be deflected
by forward machine elements. This effort will require as well a close collaboration with the eRHIC machine design efforts to ensure the best
possible forward detector acceptance.

\end{document}